\documentclass[pra,nofootinbib]{revtex4}

\pdfoutput=1

\usepackage{amsmath,amssymb}
\usepackage{epsfig}
\usepackage{color}
\newcommand{\n}{\nonumber}
\newcommand{\be}{\begin{equation}}
\newcommand{\ee}{\end{equation}}
\newcommand{\bea}{\begin{eqnarray}}
\newcommand{\eea}{\end{eqnarray}}

\begin{document}

\title{Interpolating supersymmetric pair of Fokker-Planck equations}
\author{Choon-Lin Ho}
\address{Department of Physics, Tamkang University,
Tamsui 25137, Taiwan}



\begin{abstract}

We consider Fokker-Planck equations that interpolate a pair of supersymmetrically related Fokker-Planck equations with constant coefficients.   Based on the interesting property of shape-invariance, various one-parameter interpolations of the solutions of the supersymmetric pair of Fokker-Planck systems can be directly constructed.

\end{abstract}



 \maketitle 

\section{Introduction}

One of the basic tools widely employed to study the effecst of fluctuations in
macroscopic systems is the Fokker-Planck equation (FPE) \cite{R,F}. 
Owing to  its broad applicability, it is of
great interest to obtain solutions of the FPE for various physical
situations. 

However, as for any equation in science, it is generally not easy to find analytic solutions of FPEs.  
In most cases, one can only solve the equation
approximately, or numerically. 
Nevertheless, methods of solving the FEP  have been developed in the past.
These include a change of variables,  eigenfunction expansion, variational approach, perturbation
expansion,  Green's
function, moment method, path integral, the continued-fraction
method, etc. \cite{R,F}.
Lie symmetry methods 
\cite{SS} and similarity method \cite{Ho1,Ho2,Ho3} have also been considered in solving and classifying the FPE. 

Needless to say, any method to obtain exactly solvable FPE is always welcome.  Some recent work on various aspects of solution of the FPE can be found in \cite{CCR,DM,KSF,LM,Ho4,WH,OK}. 

It is an interesting fact that 
the one-dimensional FPE with constant drift and diffusion coefficients can be transformed into a corresponding Schr\"odinger equation.  As such, any method that solves a Schr\"odinger equation exactly can be carried over to solving the FPE.  An important such method is the Darboux transformation (the time-independent
version is more commonly known as the supersymmetric (SUSY)  method in physics literature) \cite{susy}. 
This method has been employed to enlarge the number of solvable FPEs \cite{BB,Ros,SMJ,SR,IN,Ho5}.
Recently, we have further considered FPEs which are related by Darboux transformation through their corresponding Schr\"odinger equations. 
Under appropriate conditions, we have studied how an exactly solvable FPE can be obtained by the SUSY transformation from the solutions of a known FPE.  These two FPEs we shall call the SUSY FPE partners \cite{Ho5}. 

In this note, we would like to present  a simple way to generate exactly solvable FPE with solutions that interpolate those of two SUSY partners of FPEs. They form a one-parameter family of deformed FPEs between the pair of SUSY FPEs. 


\section{Fokker-Planck and Schr\"odinger equations}


The FPE is
\be
\frac{\partial}{\partial t} P(x,t; \mathbf{a}_0)=\left[-\frac{\partial}{\partial x} D(x;\mathbf{a}_0) +
\frac{\partial^2}{\partial x^2}\right]\,P(x,t; \mathbf{a}_0).\label{FPE}
\ee
The function  $P(x,t;\mathbf{a}_0)$ describes the distribution of particles in a system.  $\mathbf a_0$ is a set of parameters characterizing the system. \
$D (x;\mathbf{a}_0)$ is the drift coefficient that represents the external force acting on the particle. The constant coefficient of the second derivative term in (\ref{FPE}), which represents diffusion effect, is set to unity without loss of generality.
 The drift coefficient can be given by a  drift potential $W(x; \mathbf{a}_0)$ as  $D (x; \mathbf{a}_0)=-2 W^\prime (x; \mathbf{a}_0)$, where the prime denotes the derivative with respect to $x$.  The function $W(x; \mathbf{a}_0)$ is called the prepotential because, as will be shown below, it determines the potential of a Schr\"odinger equation related to the FPE.  As we will encounter various functions involving  parameters other than $\mathbf{a}_0$, we find it convenient to use subscript/superscript  to indicate the parameters involved, so we write  $P_0(x,t)\equiv  P(x, t;\mathbf{a}_0)$,  $D_0(x)\equiv  D (x;\mathbf{a}_0)$ and  $W_0(x)\equiv   W(x;\mathbf{a}_0)$.
 
By setting 
\be
P_0(x, t)=e^{-W_0(x)}\Psi(x,t), ~~\Psi(x,t)=e^{-\lambda(\mathbf{a}_0)\,t}\phi (x; \mathbf{a}_0),
\label{P}
\ee
we  can recast (\ref{FPE}) into
\be
H^{(0)}\phi=\lambda^{(0)} \phi^{(0)},
\label{SE}
\ee
where 
\be
H^{(0)}\equiv-\frac{\partial^2}{\partial x^2}+W_0^{\prime 2} - W_0^{\prime\prime},\label{H0}
\ee
$\lambda^{(0)}\equiv \lambda(\mathbf{a}_0)$ and $ \phi^{(0)}(x)\equiv \phi (x; \mathbf{a}_0)$. In these cases  we  use superscript to indicate the parameter involved.
The function $\phi^{(0)}$ satisfies the stationary Schr\"odinger equation with the Hamiltonian $H^{(0)}$ and eigenvalue $\lambda^{(0)}$ (for clarity of presentation, we will not indicate the independent variables and parameters if no confusion arises).
It is clear that $W_0$ defines the potential of the Schr\"odinger equation and the zero-mode $\phi^{(0)}_0\equiv \exp(-W_0)$ : $H^{(0)}\phi^{(0)}_0=0$. 

So every FPE (\ref{FPE}) with constant diffusion can be associated with a Schr\"odinger equation (\ref{SE}).
As such every solution of the Schr\"odinger equation gives a solution of the FPE. 
Suppose all the normalized eigenfunctions $\phi^{(0)}_n\equiv\phi_n(x; \mathbf{a}_0)$
($n=0,1,2,\ldots$) of $H^{(0)}$ with eigenvalues $\lambda^{(0)}_n\equiv \lambda_n(\mathbf{a}_0)$ ($\lambda^{(0)}_0=0$) are solved, then by (\ref{P})
the general solution of the FPE is 
\bea
P_0(x,t)&=&\phi^{(0)}_0(x)\,\Psi^{(0)}(x,t),\label{sol}\\
\Psi^{(0)}(x,t) &=& \sum_n c_n \phi^{(0)}_n(x)\exp(-\lambda^{(0)}_n\,t).
\n
\eea
The constant coefficients $c_n$ ($n=0,1,\ldots$) are determined from the initial profile $P_0(x, 0)=\phi^{(0)}_0(x)\sum_n c_n \phi^{(0)}_n(x)$ by
\begin{eqnarray}
c_n=\int_{-\infty}^\infty  \phi^{(0)}_n(x)\left( \phi^{(0)-1}_0(x)P_0(x,0)
\right)dx.
\end{eqnarray}
The stationary distribution as $t\to \infty$ is $P_0(x)=\phi_0^2=\exp(-2W_0)$.  If $P_0(x,t)$ is normalized, i.e., $\int \,P_0(x,t) dx=1$, then $c_0=1$.

\section{Supersymmetric pair of  FPEs}

The connection between FPE and the Schr\"odinger equation allows one to generate a Fokker-Planck system from a known one by using the supersymmetric structure of the Schr\"odinger equation \cite{Ho4}.  Note that $H^{(0)}$ is factorizable, $H^{(0)}=A^\dagger\,A$, where $A=\partial_x +W_0^\prime$ and $A^\dagger=-\partial_x +W_0^\prime$. The supersymmetric partner (Darboux transformed)  Hamiltonian is given by 
\be
\widetilde{H}=A A^\dagger =-\frac{\partial^2}{\partial x^2} + W_0^{\prime 2} + W_0^{\prime\prime}.
\label{H+}
\ee
The eigenvalues and the normalized eigenfunctions of the corresponding Schr\"odinger equation
\be
\widetilde{H}\tilde{\phi}_n=\tilde{\lambda}_n\tilde{\phi}_n, ~~n=0,1,2,\ldots
\label{SE_susy}
\ee
are related to  those of (\ref{SE}) by 
\be
{\tilde\lambda}_n = \lambda^{(0)}_{n+1},\ \ \ \ \ {\tilde\phi}_n=\frac{1}{\sqrt{\lambda^{(0)}_{n+1}}}A\,\phi^{(0)}_{n+1} ,~~~ n=0,1,2,\ldots.
\label{link}
\ee
$\widetilde{H}$ has the same spectrum as that of $H^{(0)}$ except $\lambda^{(0)}_0=0$ since $A_0\phi_0=0$. The ground state of the SUSY partner has the same energy $\lambda^{(0)}_1$ as the first excited state of the original system.

The SUSY Hamiltonian $\widetilde{H}$ does not allows us to associate it with a FPE because of the ``plus" sign between the two prepotential terms in (\ref{H+}). 
However, this is made possible if the potentials in the two Hamiltonians are shape-invariant. 
 Shape invariance means that the two potentials are similar in shape and differ only in the parameters appearing in them \cite{susy}.
 It turns out to be a  sufficient condition for exact-solvability, and it  is amazing that most of the known one-dimensional exactly solvable quantal systems possess this property. Mathematically shape invariance means the condition
\be
W^\prime (x; \mathbf{a}_0)^2 + W^{\prime\prime}(x; \mathbf{a}_0)
= W^\prime (x; \mathbf{a}_1)^2 - W^{\prime\prime}(x; \mathbf{a}_1) + R(\mathbf{a}_0).
\label{SI}
\ee
Here $\mathbf{a}_1$ is a function of $\mathbf{a}_0$ and $R(\mathbf{a}_0)$ is an $x$-independent function of $\mathbf{a}_0$.
It turns out that for  all the well-known SUSY one-dimensional solvable potentials,  $\mathbf{a}_1$ differs from $\mathbf{a}_0$ only by constant shifts, i.e., $\mathbf{a}_1=\mathbf{a}_0+\boldsymbol{\delta}$. 
The condition (\ref{SI}) then gives the eigenvalues as  \cite{susy}
\be
\lambda_n^{(0)}=\lambda_n(\mathbf{a}_0)=\sum_{k=0}^{n-1} R(\mathbf{a}_0+k \boldsymbol{\delta}).
\label{eigen}
\ee

The presence of shape invariance permits us to associate a FPE with the Hamiltonian
\be
H^{(1)}\equiv W_1^{\prime 2}-W_1^{\prime\prime},
\ee
where $W_1(x)\equiv  W_0 (x; \mathbf{a}_1)$.
It is obvious that the eigenvalues $\lambda_n^{(1)}$ and the normalized eigenfunctions $\phi_n^{(1)}(x)$ of $H^{(1)}$ are given by
\be
\lambda^{(1)}_n = \lambda_n(\mathbf{a}_1),\ \ \   \phi_n^{(1)}=\frac{\alpha}{\sqrt{\lambda^{(0)}_{n+1}}}A\,\phi^{(0)}_{n+1} ,~~~ n=0,1,2,\ldots, ~~\alpha =\pm 1.
\label{link2}
\ee
The first relation in (\ref{link2}) is obtained as follows. From (\ref{link}) and (\ref{eigen}) we have
\bea
\lambda^{(1)}_n 
&=& \tilde\lambda_n - R(\mathbf{a}_0)\n\\
&=&\sum_{k=1}^{n} R(\mathbf{a}_0+k \boldsymbol{\delta})\label{lam}\\
&=& \sum_{k=0}^{n-1} R(\mathbf{a}_1+k \boldsymbol{\delta})=\lambda_n(\mathbf{a}_1).\n
\eea

The FPE corresponding to $H^{(1)}$ has drift coefficient $D^{(1)}(x)=-2W_1^\prime$.  The two FPEs with drift coefficients related by $\mathbf{a}_0$ and $\mathbf{a}_1$ are called the supersymmetric pair FPEs.  The phase $\alpha=\pm 1$ depends on the SUSY systems.

The general solution of the partner FPE is
\bea
P_1(x,t)&=&e^{-W_1}\,\Psi^{(1)}(x,t),\\
\Psi^{(1)}(x,t) &=& \sum_n d_n \phi^{(1)}_n \exp(-\lambda^{(1)}_n t).
\n
\eea
But in order for $P_1(x,t)$ to be SUSY-related to the solution $P_0(x,t)$ (\ref{sol}) of the original FPE, we require
$\Psi^{(1)}(x,t)\sim A \Psi^{(0)}(x,t)$, i.e., we want  $\Psi^{(1)}(x,t)$ to be the SUSY transform of $\Psi^{(0)}(x,t)$.
Now we have
\bea
A \Psi^{(0)}(x,t) &=& A_0\left(\phi^{(0)-1}_0 P_0(x,t)\right) =\sum_ {n=0} ^\infty\,c_n\left(A\phi^{(0)}_n\right)\exp(-\lambda^{(0)}_n t)\n\\
&=& e^{-R(\mathbf{a}_0) t}  \sum_ {n=0} ^\infty\,c_{n+1}\alpha \sqrt{\lambda^{(0)}_{n+1}} \phi^{(1)}_n \exp(-\lambda^{(1)}_n t).
\label{AP}
\eea
Hence we can take
\bea
d_n &=&\alpha \sqrt{\lambda^{(0)}_{n+1}} c_{n+1},\\
\Psi^{(1)}(x,t) &=& e^{R(\mathbf{a}_0) t}  A \Psi^{(0)}(x,t). \n
\eea
This gives the SUSY partner solution of  $P_0(x,t)$
\bea
P_1(x,t) &=& e^{-(W_1 - R(\mathbf{a}_0) t)} \left(\partial_x +W_0^\prime\right) \left(e^{W_0}\, P_0(x,t)\right)\n\\
&=& \phi^{(1)} _0 \sum_ {n=0} ^\infty\,\alpha \sqrt{\lambda^{(0)}_{n+1}} c_{n+1}\phi^{(1)}_n \exp(-\lambda^{(1)}_n t).
\label{p1}
\eea

\section{Interpolating SUSY pair of FPEs}

Using the above relation between a pair of SUSY FPEs with shape invariance, we can generate a one-parameter family of FPEs that interpolates between the two FPE SUSY pair.  

Take a parameter $s\in [0,1]$. Suppose we can construct a set of $s$-dependent parameters $\mathbf{a}_s$ such that 
\be
\mathbf{a}_s=\mathbf{a}_0  {\rm\ for} ~~ s=0; ~~~ \mathbf{a}_s=\mathbf{a}_1  {\rm\ for}~~ s=1.
\label{a}
\ee
Then we can associate with each $s$ a prepotential  $W_s(x)\equiv W(x; \mathbf{a}_s)$ so that $W_s(x)=W_0(x)$ for $s=0$, and $W_s(x)=W_1(x)$  for $s=1$. The FPE interpolating the SUSY pair of FPEs is defined by the drift coefficient
$D_s (x)=-2 W_s^\prime (x)$.  Let us call this FPE the $s$-defomed FPE of the SUSY pair of FPEs.

Now comes a simple observation that allows us to construct solutions of the $s$-deformed FPE that interpolates those of the  two SUSY-related FPEs defined by $W_0$ and $W_1$.   
It is clear that the Schr\"odinger equation associated with the $s$-deformed FPE has 
 eigenfunctions $\phi^{(s)}_n(x)\equiv \phi_n(x; \mathbf{a}_s)$ with  eigenvalues $\lambda_n^{(s)}\equiv \lambda_n (\mathbf{a}_s)$.  Then by (\ref{lam}), we have
 \bea
 \phi^{(s)}_n(x &)&=\phi^{(0)}_n(x), ~ \lambda^{(s)}_n(x)=\lambda^{(0)}_n(x)~~{\rm when}~~s=0,\\
  \phi^{(s)}_n(x &)&=\phi^{(1)}_n(x), ~ \lambda^{(s)}_n(x)=\lambda^{(1)}_n(x)~~{\rm when}~~s=1\n .
  \eea
The normalized solution interpolating the solutions of the SUSY pair is then given by
\be
P_s (x,t)=N \phi^{(s)}_0 \sum_ {n=0} ^\infty\,\left[(1-s)c_n + s\alpha \sqrt{\lambda^{(0)}_{n+1} }c_{n+1}\right] \phi^{(s)}_n \exp(-\lambda^{(s)}_n t),
\label{Ps}
\ee
where the noramlizing constant is
\be
N=\left[(1-s)c_0 + s\alpha \sqrt{\lambda^{(0)}_{n+1} }c_1\right]^{-1}.
\ee

The above construction is valid for any $\mathbf{a}_s$ as a function of $s$ satisfying (\ref{a}).
There could be many ways to construct such $\mathbf{a}_s$.  For illustration purpose, 
 in the next section, we present two examples with  the simplest choice of $\mathbf{a}_s$. 
As mentioned before, for all the well-known one-dimensional exactly solvable shape-invariant quantum systems possessing, the relevant parameters of the two SUSY partner potentials are simply related by constant shifts, i.e. $\mathbf{a}_1=\mathbf{a}_0+ \boldsymbol{\delta}$ where $\boldsymbol{\delta}$ is a set of constants.  Thus the simplest choice of $\mathbf{a}_s$ is
\be
\mathbf{a}_s=\mathbf{a}_0+ s \boldsymbol{\delta}.
\label{choice}
\ee
As shown in [21],  this choice appears naturally in most of the one-dimensional solvable quantum models, if one interpolates the SUSY pair of  the Hamiltonians by their transformed versions, i.e.,  
\be
(1-s) \left[e^{W_0}H_0 e^{-W_0}\right]  + s \left[e^{W_1}H_1 e^{-W_1}\right]. 
\label{C1}
 \ee
 This requirement, which involves the Hamiltonians $H_0$ and $H_1$,  is more restrictive than (\ref{a}), which only involves the parameters $\mathbf{a}_0$ and $\mathbf{a}_1$.  The construction presented here is valid for general $\mathbf{a}_2$.

\section{Examples}

As illustration, we  construct the one-parameter deformed families of two examples of SUSY FPE pairs, with the dirft potential related to the radial oscillator and the Morse potential.

\subsection*{The radial oscillator}

The prepotential for the radial oscillator potential is $W_0(x; \mathbf{a}_0)=\omega x^2/4- (\ell+1)\ln x$, where $x\in [0,\infty)$ and $\omega,  \ell>0$.  In this case  $\mathbf{a}_0=(\omega, \ell)$ and $\boldsymbol{\delta}=(0, 1)$, i.e., $\mathbf{a}_1=(\omega, \ell+1)$ \cite{susy}.  As only the parameter $\ell$ is shifted in the SUSY transformation, we shall use its values to label the two SUSY pair. 

The wavefunctions are
\be
\phi^{(\ell)}_n(x)\equiv \phi_n (x; \mathbf{a}_0)=N_{n\ell} y^{\frac{\ell+1}{2}} e^{-\frac{y}{2}} L_n^{\ell+\frac12}(y), ~~ y\equiv \frac12 \omega x^2,
\ee
The normalization constant is
\be
N_{n\ell}= (2\omega)^\frac14 \left[\left( \begin{array}{c}
 n+\ell +\frac12\\
n
 \end{array} \right) \Gamma(\ell+\frac32)
 \right]^{-1},
\ee
where the two factors in the square-bracket are the binomial coefficient and the Gamma function, respectively. 
Eigenvalues are given by $\lambda_n^{(0)}=2n\omega, ~~n=0,1,2,\ldots$.

It is easy to check that
\be
A \phi^{(\ell)}_{n+1}(x)=-\sqrt{\lambda^{(\ell)}_{n+1}}\phi^{(\ell+1)}_n(x),
\ee
so the factor $\alpha$ in (\ref{link2}) is $\alpha=-1$ in this case.

The choice (\ref{choice}) leads  to the interpolating prepotential $W_s(x)=W(x; \mathbf{a}_s)$ with $\mathbf{a}_s=(\omega, \ell+s)$.

Fig.\,1 gives the plots of the radial oscillator drift potential $W_s(x)$ and several  normalized $P_s (x,t)$ for different $s$ with $\omega=\ell=1$.  For the graphs of $P_s(x,t)$, we have scaled down the prepotential  $W_0(x)$ 3 times for a better visual comparison with $P_s(x,t)$.   The coefficients $c_n$ used in (\ref{Ps}) are $c_0=5, c_1=c_2=1$ and 
$c_n=0 (n\geq 3)$.

\subsection*{The Morse potential}

In this case the drift potential is
$W_0(x; \mathbf{a}_0)=\alpha x + \beta e^{-x}, ~~\alpha, \beta> 0, ~~\mathbf{a}_0=(\alpha, \beta)$.   Under SUSY transformation, $\mathbf{a}_1=(\alpha-1, \beta)$ \cite{susy}.  Similar to the precious example, here we shall use the only changing parameter $\alpha$  to label the two SUSY pair. 

Wavefunctions are given by: 
\be
\phi^{(\alpha)}_n(x)\equiv \phi_n (x; \mathbf{a}_0)=N_{n\alpha} y^{\alpha-n} e^{-\frac{y}{2}} L_n^{2(\alpha-n)}(y), ~~ y\equiv 2\beta e^{-x},
\ee
with normalization constant
\be
N_{n\alpha}=\left[ \frac{2(\alpha-n)\Gamma(n+1)}{ \Gamma(2\alpha-n+1)} \right]^{\frac12}.
\ee
and the corresponding eigenvalues $\lambda_n^{(0)}=\alpha^2-(\alpha-n)^2, ~~n=0,1,2,\ldots, {\rm\ largest\  integer} <\alpha$.
It is easy to check that
 the factor $\alpha$ in (\ref{link2}) is $\alpha=1$ in this case.
The choice (\ref{choice}) gives $\mathbf{a}_s=(\alpha-s, \beta)$.

In Fig.\,2 we plot the Morse drift potential  $W_s(x)$ and $P_s(x,t)$  for  $\alpha=5, \beta=1$, $c_0=3, c_1=2, c_2=1$ and $c_n=0 (n\geq 3)$.  Again for easy  visual comparison, we have plotted $(W_0)+3)/4$ instead of $W_0$ in the graphs of $P_s (x,t)$. . 

From both Fig.\,1 and 2, it is seen that the function $P_s(x,t)$ is confined within the potential well $W_s(x)$. For small times, $P_s(x,t)$ could look very different for different values of $s$. But for large times, they become quite similar.  This is expected as $t\to \infty$, only the $n=0$ term in $P_s(x,t)$ remains as $\lambda_0^{(s)}=0$, i.e., $P_s(x,t)\sim \phi_0^{(s)2}$, which are similar in shape.

\section{Summary}

We have presented a simple construction of one-parameter deformed FPEs with solutions that interpolate a pair of 
SUSY FPEs.  The construction is very general: it only requires the condition of shape-invariance,  which amazingly
is satisfied by most of the well-known exactly solvable one-dimensional quantum systems, and the interpolating condition of the parameters related by the shape-invariance, namely, Eq.\,(\ref{a}).   

We have demonstrated the procedure by two examples using the simplest interpolation rule, the linear shift (\ref{choice}). 
Other deformations are possible, as long as the rule  (\ref{a}) is satisfied.
For instance, following [21], one can consider simply interpolating  the two SUSYHamiltonians,
 \be
(1-s) H_0   + s H_1
 \ee
 instead of the gauge-transformed Hamiltonians as in (\ref{C1}).
In this case the parameters  $\mathbf{a}_s$ are mostly nonlinear in $s$ \cite{OPS}.  
In this case the parameters  $\mathbf{a}_s$ are mostly nonlinear in $s$ \cite{OPS}.  
For example, in the radial oscillator case, the parameter $\ell$ is   deformed to $[\sqrt{4(\ell+1)(\ell+2s)}-1]/2$.
Nonetheless, our construction is applicable.

\section*{Acknowledgments}
The work is supported in part by the Ministry of Science and Technology (MoST)
of the Republic of China under Grant   NSTC 112-2112-M-032-007.


\begin{figure}[ht] \centering
\includegraphics*[width=8cm,height=6cm]{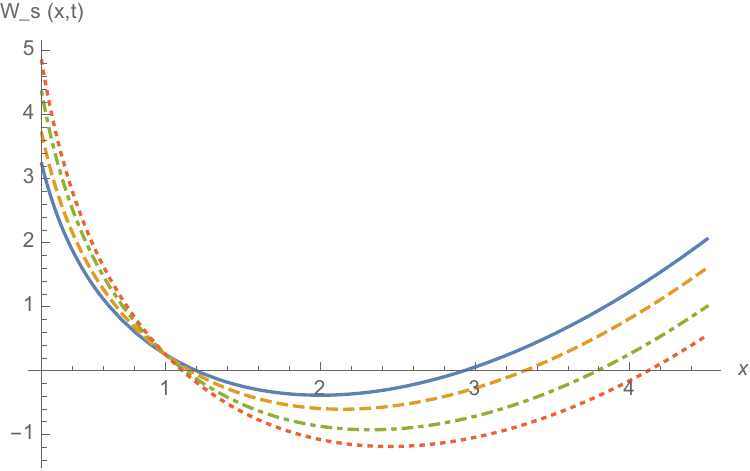}\hspace{1cm}
\includegraphics*[width=8cm,height=6cm]{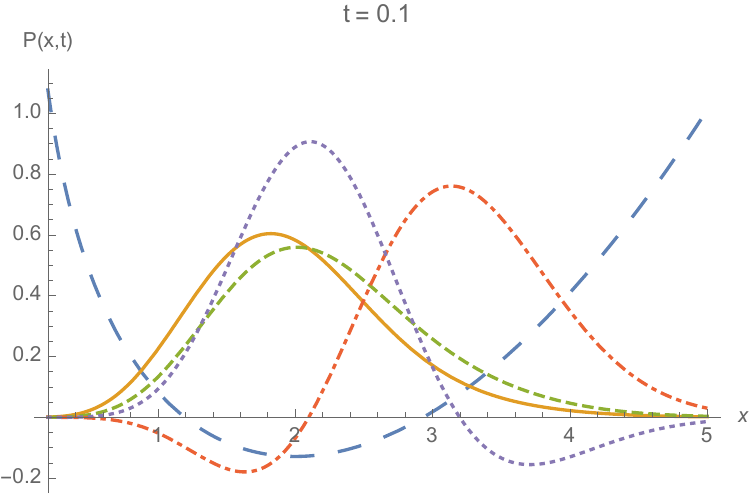}\\
\includegraphics*[width=8cm,height=6cm]{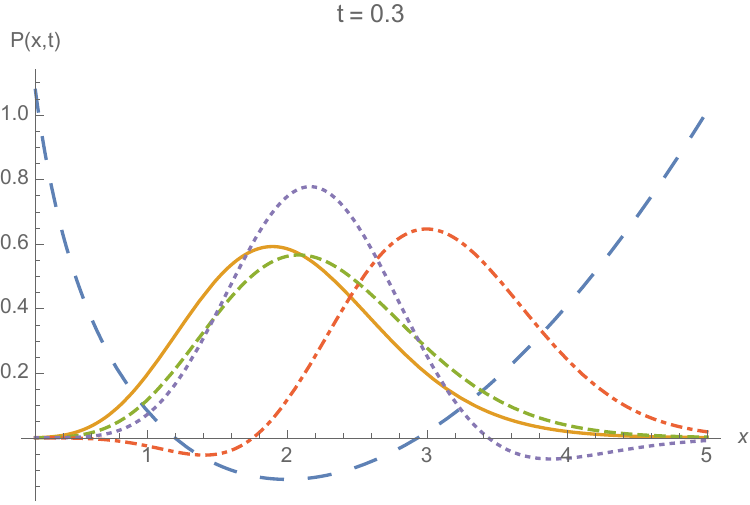}\hspace{1cm}
\includegraphics*[width=8cm,height=6cm]{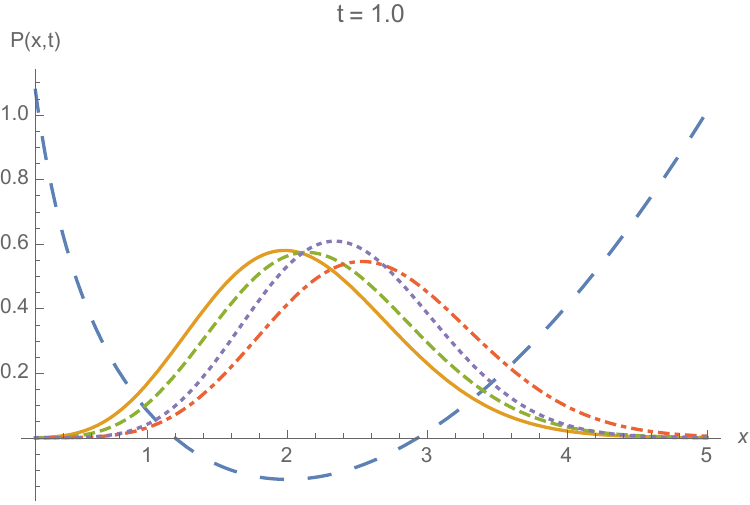}
\caption{The first plot presents graphs of $W_s (x,t)$ of the radial oscillator. The next three plots show the scaled radial oscillator drift potential $W_0(x)/3$ (large dashed line) and the normalized $P_s(x,t)$ versus $x$ with $\omega=\ell=1$, $c_0=5, c_1=c_2=1$ and $c_n=0 (n\geq 3)$, time $t=0.1$, $0.03$, $0.5$, $1.0$. The deformed parameter $s$ are $s=0$ (solid), $0.3$ (dashed), $0,7$ (dotdashed), and $1.0$ (dotted).} 
\end{figure}


\begin{figure}[ht] \centering
\includegraphics*[width=8cm,height=6cm]{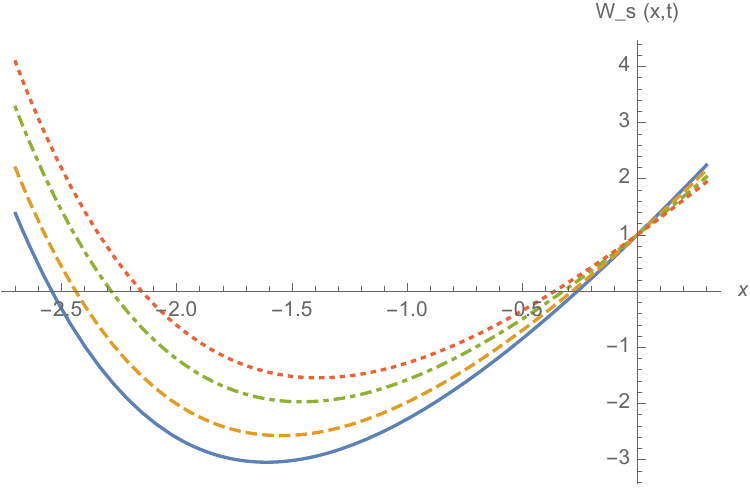}\hspace{1cm}
\includegraphics*[width=8cm,height=6cm]{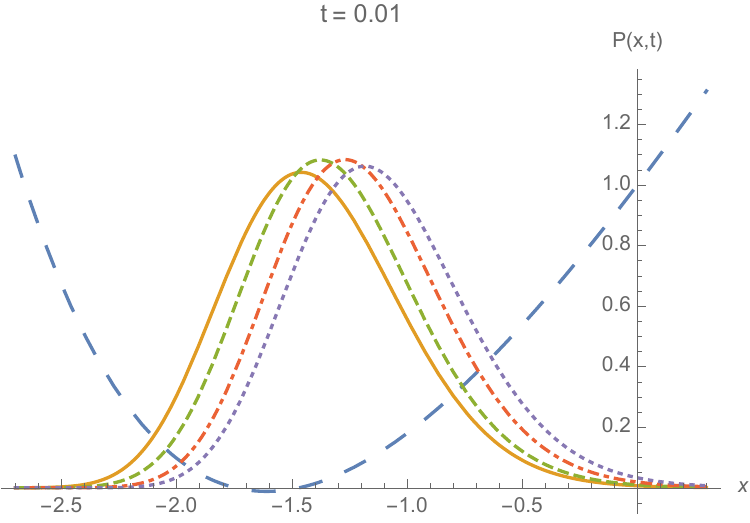}\\
\includegraphics*[width=8cm,height=6cm]{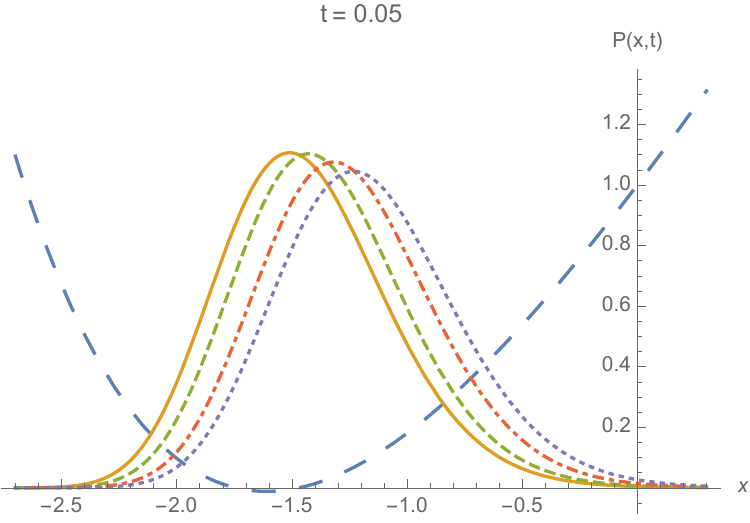}\hspace{1cm}
\includegraphics*[width=8cm,height=6cm]{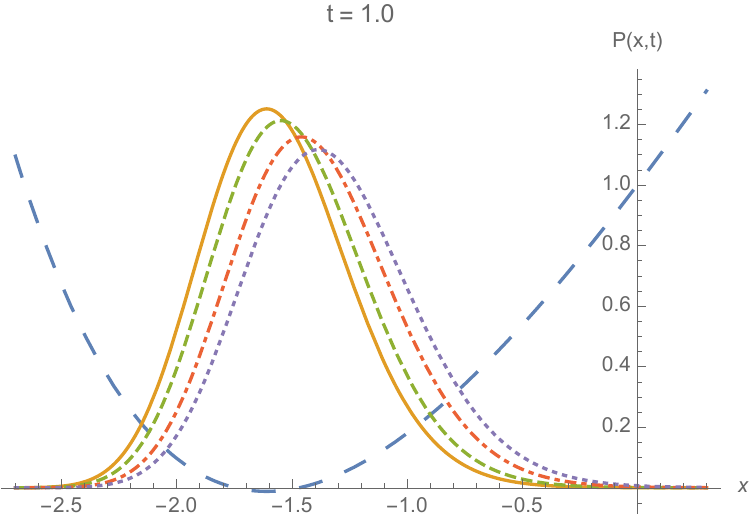}

\caption{The first plot presents the graphs of $W_s (x,t)$ of the Morse potential. The next three plots show the scaled Morse drift potential $(W_0(x)+3)/4$ (large dashed line) and $P_s(x,t)$ versus $x$ for $\alpha=5, \beta=1$, $c_0=3, c_1=2, c_2=1$ and $c_n=0 (n\geq 3)$ and time $t=0.01$, $0.05$, $1.0$, $5.0$. The deformed parameter $s$ are $s=0$ (solid), $0.3$ (dashed), $0,7$ (dotdashed), and $1.0$ (dotted).} 
\end{figure}

\end{document}